\documentclass[preprint]{aastex61}

\newcommand\aastex{AAS\TeX}

\submitjournal{ApJ}

\shorttitle{\aastex\ Hierarchical cluster assembly}
\shortauthors{Dalessandro et al.}

\begin{document}

\title{First phase space portrait of a hierarchical stellar structure in the Milky Way}

\correspondingauthor{Emanuele Dalessandro}
\email{emanuele.dalessandro@inaf.it}


\author{E. Dalessandro}
\affiliation{INAF - Astrophysics and Space Science Observatory Bologna, Via Gobetti 93/3 40129 Bologna - Italy}

\author{A.L. Varri}
\affiliation{Institute for Astronomy, University of Edinburgh, Royal Observatory, Blackford Hill, Edinburgh EH9 3HJ, UK}

\author{M. Tiongco}
\affiliation{University of Colorado, JILA and Department of Astrophysical and Planetary Sciences, 440 UCB, Boulder, CO 80309 - USA}

\author{E. Vesperini}
\affiliation{Department of Astronomy, Indiana University, Swain West, 727 E. 3rd Street, IN 47405 Bloomington - USA}

\author{C. Fanelli}
\affiliation{Dipartimento di Fisica e Astronomia,  Via Gobetti 93/2 40129 Bologna - Italy}
\affiliation{INAF - Astrophysics and Space Science Observatory Bologna, Via Gobetti 93/3 40129 Bologna - Italy}

\author{A. Mucciarelli}
\affiliation{Dipartimento di Fisica e Astronomia,  Via Gobetti 93/2 40129 Bologna - Italy}
\affiliation{INAF - Astrophysics and Space Science Observatory Bologna, Via Gobetti 93/3 40129 Bologna - Italy}

\author{L. Origlia}
\affiliation{INAF - Astrophysics and Space Science Observatory Bologna, Via Gobetti 93/3 40129 Bologna - Italy}

\author{M. Bellazzini}
\affiliation{INAF - Astrophysics and Space Science Observatory Bologna, Via Gobetti 93/3 40129 Bologna - Italy}

\author{S. Saracino}
\affiliation{Astrophysics Research Institute, Liverpool John Moores University, 146 Brownlow Hill, Liverpool L3 5RF, UK}

\author{E. Oliva}
\affiliation{INAF - Osservatorio Astrofisico di Arcetri, Largo Enrico Fermi 5, 50125 Firenze, Italy}

\author{N. Sanna}
\affiliation{INAF - Osservatorio Astrofisico di Arcetri, Largo Enrico Fermi 5, 50125 Firenze, Italy}

\author{M. Fabrizio}
\affiliation{INAF - Osservatorio Astronomico di Roma, Via Frascati 33, 00078, Monte Porzio Catone (Roma), Italy}
\affiliation{Space Science Data Center - ASI, Via del Politecnico s.n.c., 00133 Roma, Italy}

\author{A. Livernois}
\affiliation{Department of Astronomy, Indiana University, Swain West, 727 E. 3rd Street, IN 47405 Bloomington - USA}







 
\begin{abstract}
 We present the first detailed observational picture of a possible ongoing massive cluster hierarchical assembly
in the Galactic disk as revealed by the analysis of the stellar full phase-space 
(3D positions and kinematics and 
spectro-photometric properties) of an extended area ($6^{\circ}$ diameter) surrounding the well-known {\it h} and $\chi$ Persei 
double stellar cluster in the Perseus Arm.
Gaia-EDR3 shows that the area is populated by seven co-moving clusters, 
three of which were previously unknown, and by an extended and quite massive ($M\sim10^5 M_{\odot}$) halo. 
All stars and clusters 
define a complex structure with evidence of possible mutual interactions in the form of intra-cluster 
over-densities and/or bridges. They share the same chemical abundances (half-solar metallicity)
and age ($t\sim20$ Myr) within a small confidence interval and the stellar density distribution 
of the surrounding diffuse stellar halo resembles that of a cluster-like stellar 
system.  The combination of these evidences suggests 
that stars distributed within 
a few degrees from {\it h} and $\chi$ Persei are part of a common, sub-structured stellar complex 
that we named  LISCA I. 
Comparison with results obtained through direct $N$-body simulations suggest that LISCA I may be at an intermediate stage
of an ongoing cluster assembly that can eventually evolve in a relatively 
massive (a few $10^5 M_{\odot}$) stellar system. 
We argue that such cluster formation 
mechanism may be quite efficient in the Milky Way and disk-like galaxies and, as a consequence, 
it has a relevant impact on our understanding of cluster formation efficiency as a function 
of the environment and redshift.  

\end{abstract}




\section{Introduction}
It is widely accepted that most ($70-90\%$) stars in galaxies form in groups, 
clusters or hierarchies, and spend some time gravitationally bound with their siblings 
when still embedded in their progenitor molecular cloud \citep{lada03,PZ10}. 
Indeed, young stars tend to be 
grouped together into a hierarchy of scales where smaller and denser sub-regions 
(with a typical dimension of a few parsecs) are inside larger and looser ones (up to hundreds of parsecs; 
\citealt{elmegreen10}). 
The majority of such systems will be disrupted in their first few million years of existence, 
due to mechanisms possibly involving gas loss driven by stellar feedback \citep{moeckel10} or encounters with 
giant molecular clouds \citep{gieles06}. Nonetheless, a fraction of proto-clusters will survive the embedded 
phase and remain bound over longer timescales. The process of clustered star formation has major 
implications for many fundamental astrophysical areas of research including the star 
formation process itself \citep{mckee07}, the early interplay between stellar and gas dynamics and the consequences 
of gas expulsion for the cluster disruption \citep{baumgardt07}, the possible 
formation of gravitational wave sources \citep{dicarlo19}, 
and the dynamical properties of young star clusters \citep{mcmillan07}.
Cluster formation has also key implications on our understanding of the assembly process of galaxies 
in a cosmological context. Indeed, major star-forming episodes in galaxies are typically accompanied 
by significant star cluster production \citep{forbes18} and the main properties of these systems are thus strictly 
connected with those of their hosts \citep{brodie06,dalessandro12}. 
Indeed, massive star clusters may play a role in the formation 
of galactic sub-structures, and their partial or total dissolution contributes to the overall mass budget
and stellar population properties of the hosts. 

However, in order to efficiently exploit stellar clusters as tracers of galaxy and large-scale structure formation, 
it is essential to understand the physical processes setting their initial properties such as mass, 
structure and chemical composition and how they evolve across the cosmic time.
In the last decade, many observational and theoretical studies have greatly enriched our understanding of 
the formation and early evolution of star clusters (e.g.,
\citealt{goodwin04,allison10,parker14,adamo15,adamo20}). However, questions concerning the possible presence 
of unifying principles governing the formation of different stellar systems are still unanswered  
(e.g., \citealt{bonnell03,longmore14,banerjee14,banerjee15,banerjee17,kuhn19,kuhn20}). 
In addition, the presence in old globular clusters of multiple stellar populations characterized 
by specific abundance patterns in a number of light elements (see for example \citealt{bastian18,gratton19} for recent reviews) 
has indeed raised new key questions on the physical 
mechanisms at the basis of clusters' formation, and their dependence on the environment and the formation epoch \citep{krumholz19}. 
As a matter of fact, in spite of the tremendous observational and theoretical efforts 
our understanding of the star clusters' formation history and the underlying physical processes is still in its infancy.

The unprecedented precision and sensitivity in measuring stellar parallaxes and proper motions 
secured by the Gaia mission \citep{gaia18,gaia20} now provide key 
tools to study in detail nearby star-forming regions and to use them as ideal laboratories 
to shed new light on our understanding of cluster formation and early evolution. Indeed, several papers 
have been published recently on the subject (e.g. \citealt{beccari18,zari18,kuhn19,meingast19,lim20}). 

In this context, in the present study, we focus on the neighborhood of the well-known young double clusters {\it h} and $\chi$ Persei 
(NGC~884 and NGC~869, respectively). This area is located in the Galactic Perseus Arm, 
which includes massive star-forming regions W3-W4-W5, with two giant H II regions (W4 and W5), 
a massive molecular ridge with active formation (W3), and several embedded star clusters and/or associations 
\citep{carpenter00}. 
The paper is organized as follows.
The adopted data-set is presented in Sections~2, in Sections~3 and 4 the main spectro-photometric and
kinematic properties of the area are described respectively. A comparison with a set of $N$-body simulations following the
violent relaxation phase and its subsequent evolution is described in Section~5. The main conclusions are discussed in
Section~6. 

\begin{figure}
	\includegraphics[width=\columnwidth]{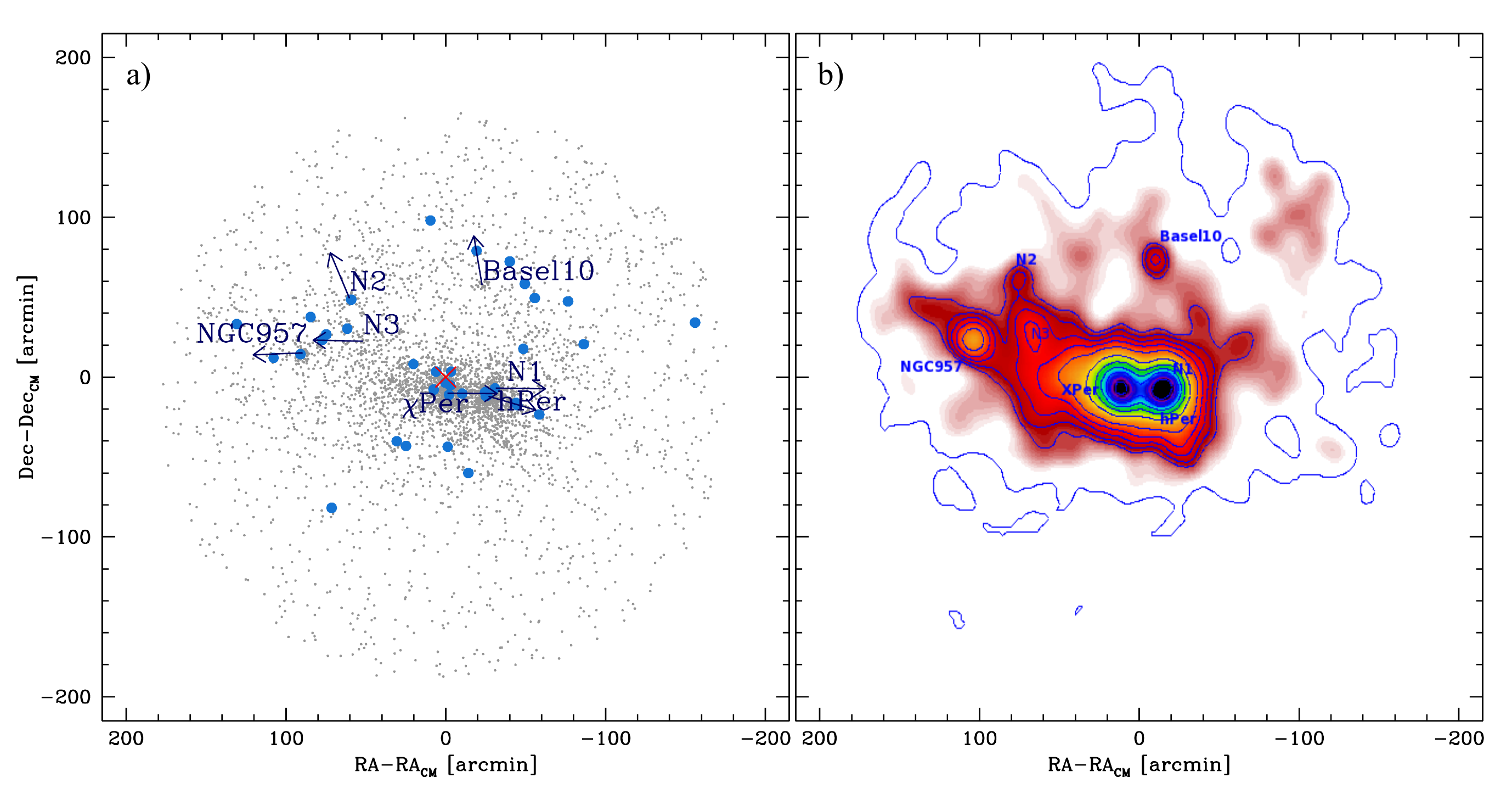}
    \caption{{\bf Panel a}: Map of the Gaia sources (G$<$16) selected in proper motion and parallax as 
    described in the text with respect to the position of the system barycenter (red cross). 
    Blue arrows indicate the mean cluster motion in both the radial and tangential components with respect 
    to the mean motion of the system. Light blue circles represent the positions of the SPA-TNG spectroscopic targets. 
    {\bf Panel b}: 2D surface density map obtained for the same stars as in the left panel. 
    The blue contour levels span from 3 to $50 \sigma$ with irregular steps.}
    \label{fig:map}
\end{figure}

\section{The Perseus Arm complex under the Gaia and SPA-TNG lenses} 
For the present work we used photometric and astrometric information obtained from 
Gaia Early Data Release 3 (EDR3), we refer the reader to \citet{gaia18,gaia20} for details about the survey and available data. 
These data were complemented 
with high-resolution optical
and near-infrared spectra obtained respectively with the HARPS-N \citep{cosentino14} 
and the GIANO-B \citep{oliva12,tozzi16} 
spectrographs mounted at the Italian Telescopio Nazionale Galileo (TNG) operated on the island of 
La Palma at the Spanish Observatorio del Roque de los Muchachos of the Instituto de Astrofisica de Canarias 
by the Fundac\'ion Galileo Galilei of the INAF (National Institute for Astrophysics). 
These observations are part of the TNG Large Program titled {\it SPA - Stellar Population Astrophysics: 
the detailed, age-resolved chemistry of the Milky Way disk} (Program ID A37TAC13, PI: L. Origlia), 
aimed at measuring detailed chemical abundances and radial velocities of the luminous stellar 
populations of the Milky Way (MW) thin disk and its associated star clusters \citep{origlia19}. 
Within SPA we collected 
HARPS-N spectra at $R\sim115,000$ in the 378 - 691 nm range and GIANO-B spectra at $R\sim50,000$ 
in the 950 - 2450 nm range for 37 stars among blue and red 
supergiants and young Main Sequence stars in the analysed area. 
HARPS-N spectra are automatically processed by the instrument data reduction software pipeline.
GIANO-B spectra have been reduced using the \texttt{GOFIO} data reduction software code \citep{rainer18}.

\subsection{Area characterization and cluster identification}
We selected stars in the Gaia-EDR3 catalog, distributed within a circular projected 
area with radius of 3 degrees from {\it h} and $\chi$ Persei (Figure 1), located at $1 \sigma$ from their mean 
distance ($d=2.353_{-0.197}^{+0.178}$ kpc, i.e. those stars having parallaxes in the range $0.385<\mu<0.465$ mas) 
and moving with similar mean velocities as the two clusters ($\mu_{\alpha} cos\delta=-0.64$ mas/yr and $\mu_{\delta} 
= 1.17$ mas/yr)
within a very strict tolerance range of $\pm0.4$mas/yr, corresponding to about 4 km/s at their distance.

As shown in Figure 1a, in addition to {\it h} and $\chi$ Persei, other clusters, clumps and elongated structures 
are clearly visible in the area after applying the described proper motion and parallax selections. 
Two of them are the known open clusters NGC~957 and Basel~10.  
To identify the additional cluster-like components, we used the clustering 
algorithm \texttt{DBSCAN} (Density-Based Spatial Clustering of Applications with Noise).  
\texttt{DBSCAN} is a density-based clustering algorithm that interprets clusters as regions of high density 
separated by areas of low density in space, without requiring any additional prior. 
Only two parameters need to be set in the algorithm, which are $eps$ and $min_{samples}$. 
$eps$ defines the radius of neighborhood around a point x, and $min_{samples}$ represents 
the minimum number of neighbors within the $eps$ radius. Higher $min_{samples}$ or lower $eps$ values 
indicate higher densities necessary to identify a sub-system as a cluster. 
Inspecting Figure 1, it is quite evident that clusters in the considered 
region have different sizes and numbers of members, 
and therefore different densities. 
For this reason, the clustering analysis needs to be run more than once.  
We first focused on the central region, where {\it h} and $\chi$ Persei are located. 
We applied \texttt{DBSCAN} on a sub-sample of stars having magnitude G $< 16$ and
by imposing $min_{samples}$ = 45 and $eps = 0.04$. 
This choice allowed us to isolate the main structures near the center 
of the map and to identify three main groups. 
Two of them are obviously {\it h} and $\chi$ Persei, the most dense and easy to identify, 
but a third component, very close to {\it h} Persei - that we will call hereafter N1 - has been also identified. 
Its location is highlighted in Figure~2. Then, we moved to the more external part, 
where less dense structures are present. We applied the \texttt{DBSCAN} algorithm only 
to such stars with the same magnitude cuts and using different parameters: 
$min_{samples}$ = 41 and $eps = 0.06$ and we identified in this way four additional structures: 
the already known Basel~10
and NGC~957, and the two new additional systems that we name N~2 and N~3 (see Table~1 for a summary of the clusters' main properties).
We note that the identification of the new clusters is quite solid against different settings of the \texttt{DBSCAN} search. In
particular, we checked that results do not vary when using different $eps$ values in the range 0.03-0.09.

\begin{figure}
\centering
	\includegraphics[width=12cm]{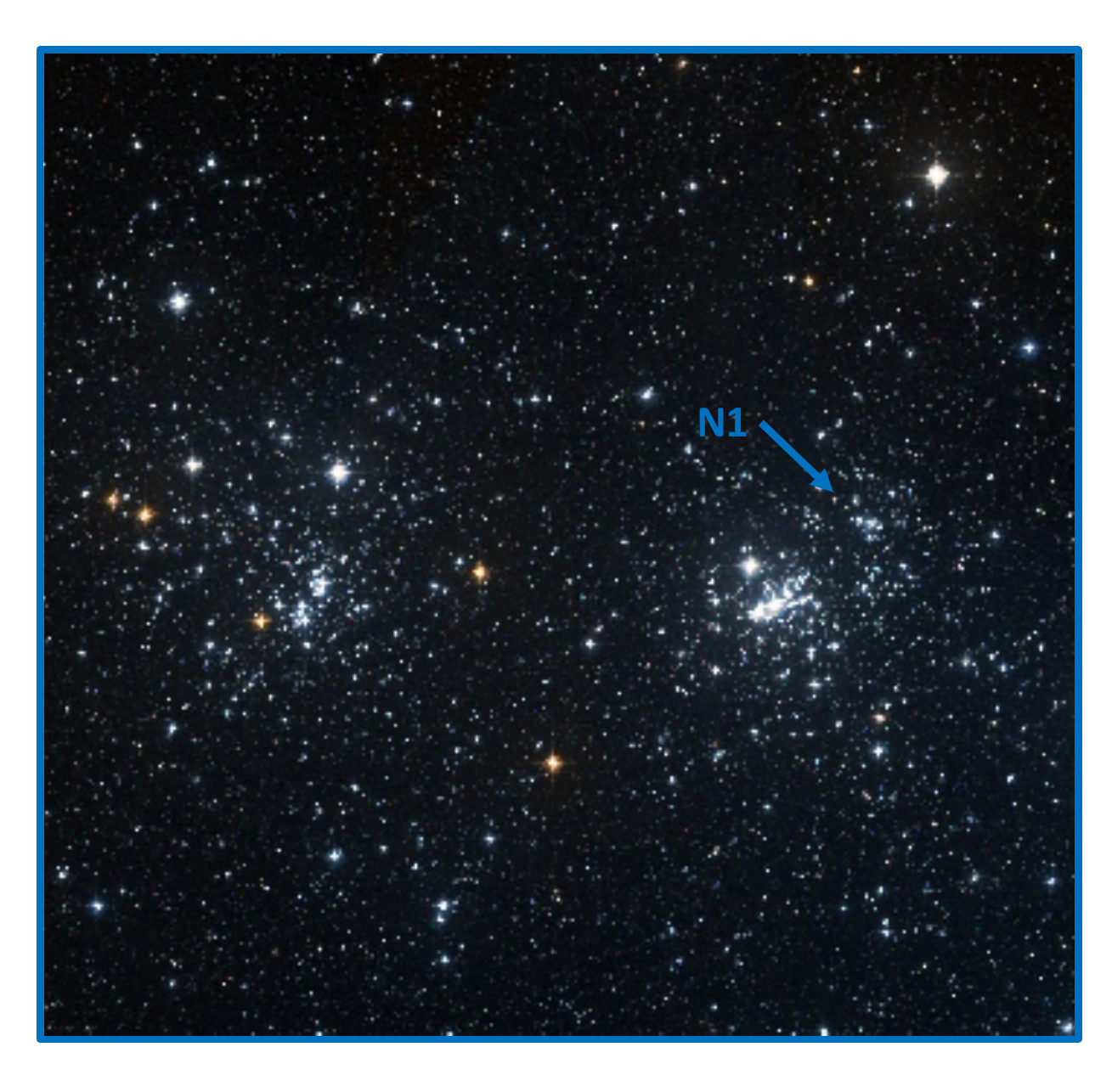}
    \caption{Zoomed view of the {\it h} and $\chi$ Persei region with highlighted the location of the newly identified system N1.}
    \label{fig:map}
\end{figure}

\begin{table*}
\begin{footnotesize}
\begin{center}

\setlength{\tabcolsep}{0.2cm}

\caption{Coordinates and mean velocities of the clusters identified in LISCA I}
\label{PHOT_PAR}
\begin{tabular}{@{}rccccc}
\hline
Cluster	  &  RA(J2000)  &   Dec(J2000)   &     $\mu_{\alpha}cos(\delta)$   &  $\mu_{\delta} $ [mas/yr] &     LOS-V  \\
           &   (hh:mm:ss)         &    (dd:mm:ss)            &      [mas/yr]         &       [mas/yr]            &     (km/s) \\
          \hline
$\chi$~Per  &  02:22:07.281 & 57:09:48.62 &  -0.621 &   -1.164 &  -45.3 \\
$h$~Per     &  02:18:58.429 & 57:07:43.01 &  -0.662 &   -1.170 &  -45.6 \\
NGC~957     &  02:33:23.532 & 57:33:17.32 &  -0.293 &   -1.120 &  -44.8 \\
Basel~10    &  02:19:23.301 & 58:17:25.58 &  -0.466 &   -0.881 &  -53.0\\
N~1         &  02:18:23.612 & 57:12:40.79 &  -0.682 &   -1.165 &  -47.1 \\
N~2         &  02:29:52.466 & 58:08:19.30 &  -0.433 &   -0.923 &  -49.7 \\
N~3         &  02:28:44.542 & 57:41:37.94 &  -0.437 &   -0.957  &  -46.2 \\
\hline

\end{tabular}

\end{center}
\end{footnotesize}
\end{table*}

The presence of these additional clusters is remarkably evident in the two-dimensional (2D) 
density map shown in Figure 1b, which has been obtained by transforming 
the distribution of star positions into a smoothed surface density function through the use of a 
gaussian kernel with width of $3\arcmin$ (see for example \citealt{dalessandro15}). 
Interestingly, all star clusters in the complex show some degree of elongation, 
which is particularly relevant in the case of N~3. In addition, it is worth noting the presence of 
a significant overdensity bridging the regions between {\it h} and $\chi$ Persei and extending 
out to NGC~957 and N~2, which may be suggestive of ongoing tidal interactions or the remnants of primordial 
filamentary structures, which are now commonly found in the Milky Way disk \citep{kounkel19}. 
Figure 1b also shows the isodensity curves as obtained from 3 to 50 $\sigma$, where $\sigma$ 
is the `background' density dispersion observed for stars located at about 3 degrees from {\it h} and $\chi$ Persei 
in the South-East quadrant. They clearly suggest the presence of a diffuse low-density stellar halo. 
The existence of a possible halo stellar distribution in the proximity of {\it h} and $\chi$ Persei
(at distances $<0.5$ deg from the two clusters) was already proposed by \citet{currie10}. 
Thanks to the Gaia astrometric information here we find that the halo actually extends with a rather irregular shape 
for at least 3 degrees from {\it h} and $\chi$ Persei at a $3 \sigma$ significance level.

{\it Are these co-moving clusters and diffuse halo (along with its sub-structures) part of a single system? 
Or are they just independent neighbors?} 
The answers to these questions can have key implications on our understanding of their formation, 
evolution and, more in general, on stellar cluster formation and early dynamical processes. 
To address them we will examine two main lines of evidence that will be detailed in Sections~3 and 4.

\begin{figure}
	\includegraphics[width=\columnwidth]{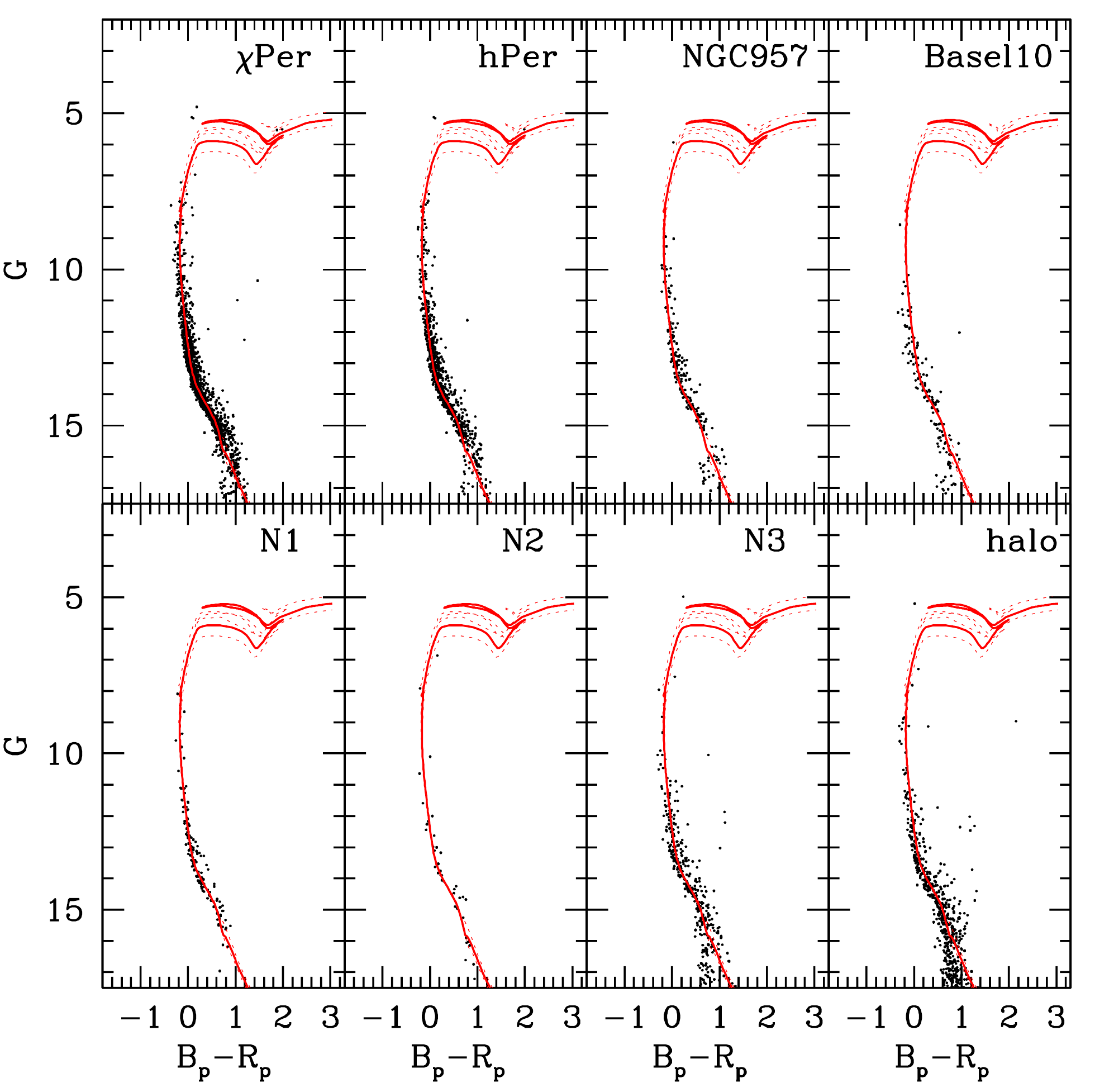}
    \caption{Gaia CMDs of the clusters detected around $h$ and $\chi$ Persei. The solid red line represents an isochrone \citep{bressan12} 
    of 20 Myr with metallicity [Fe/H]$=-0.29$, placed at a distance of 2.3 kpc and assuming $A_v=1.65$ mag.  Dashed lines are 
    isochrones at 15 Myr and 25 Myr.
   In general they reproduce reasonably well the {\it h} and $\chi$ Persei CMDs and they have been used as reference in the 
   other panels.}
    \label{fig:map}
\end{figure}

\section{Ages and chemical composition}
We first derived stellar cluster and halo star chemical abundances by performing a 
detailed analysis of the high-resolution near-infrared spectra obtained with GIANO-B 
within the SPA-TNG Large Program.
Chemical abundances and abundance patterns for all the most important metals 
are computed by using the \texttt{MARCS} model atmospheres \citep{gustafsson08} and the spectral synthesis technique 
implemented in the \texttt{Turbospectrum} code \citep{alvarez98,plez12} for the red supergiants in the sample. 
A detailed and comprehensive description of the analysis and results will be presented in a forthcoming paper (C. Fanelli et
al. in preparation). 
For the purpose of this study we report that
all the analyzed stars share the same metallicity within the uncertainties 
(which are typically of the order of 0.05 dex) with a mean value of [Fe/H]$=-0.29\pm0.03$ dex 
and about solar-scaled [$\alpha$/Fe]. 
We compared these estimates with predictions from the Besan\c{c}on Galaxy model \citep{robin03} 
for stars in this region and selected in distance and proper motions as done before, and with results
 from surveys of the galactic disk \citep{mikolaitis14,hayden14} at the galactocentric distance of 
 {\it h} and $\chi$ Persei. 
 The comparison distributions have a similar mean metallicity to that of stars analyzed here, 
 but they are significantly broader extending for about 1.5 dex in the range $-1.0<[Fe/H]<0.3$ dex and 
 thus suggesting that the lack of significant spread is not the expected outcome of the disk 
 population properties in this area.  
 
Secondly, we constrained cluster and halo ages by comparing the differential reddening corrected 
Gaia (G, Bp-Rp) color magnitude diagrams (CMDs) with a suitable set of isochrones (Figure~3). 
Extinction values have been 
assigned to each star in the Gaia catalog by interpolation with the \citet{schlegel98} extinction maps and 
corrected to estimates by \citet{schlafly11}.
An accurate age derivation is certainly a challenge for young and scarcely populated 
clusters (in particular for N1, N2 and N3) as the turn-off region is significantly affected 
by low number counts (and relative fluctuations), stellar 
rotation and binarity \citep{li19}, which can make the adoption of a direct isochrone 
fitting approach unreliable. 
However, we stress that here we are mainly interested in investigating the presence of significant age differences
among clusters.
\citet{currie10} found that {\it h}, $\chi$ Persei and their surrounding stars are coeval
with an age $t_{age}\sim14$Myr. 
We extended such an analysis by using a set of PARSEC isochrones \citep{bressan12}.  
We adopted the derived mean metallicity ([Fe/H]=-0.29) and distance ($d=2.353$ kpc) 
and an average extinction value $A_V=1.65$ mag. 
We find that all systems are compatible with being almost 
coeval with ages in the 15-25 Myr range represented by the three isochrones shown in each 
CMD of Figure 3. 
While uncertainties on the distance can 
surely impact the derived ages, we stress that the main source of uncertainty is the 
limited number of bright stars needed to anchor the best-fit isochrone.  
Interestingly we find that also halo stars share the same age as clusters in agreement with 
previous findings \citep{currie10}. Only Basel~10 is possibly slightly older ($\sim30$ Myr).

\begin{figure}
	\includegraphics[width=\columnwidth]{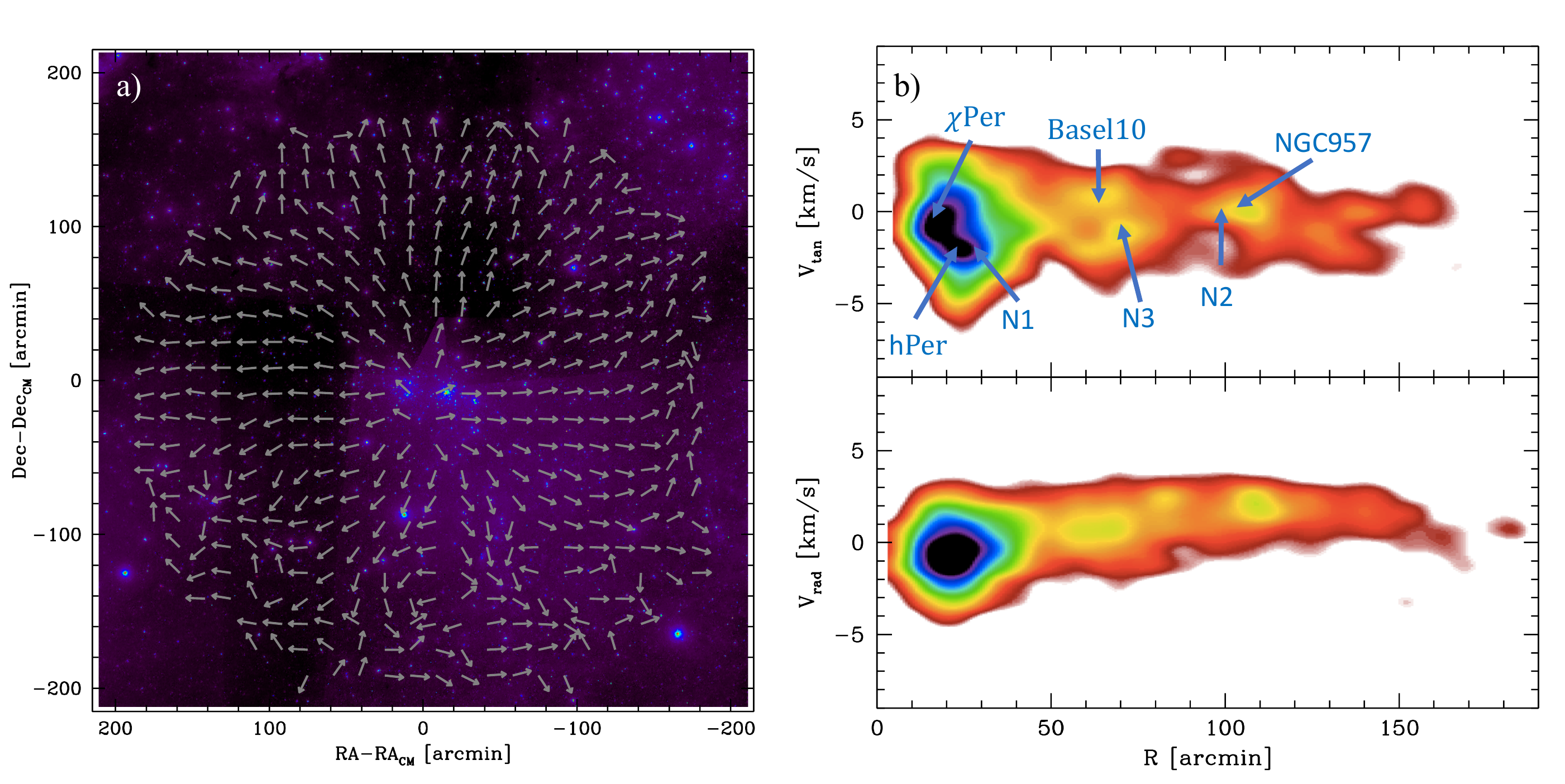}
    \caption{{\bf Panel a}: Velocity map obtained for stars with G$<16$ superimposed to a Digital Sky Survey 2 
    false color image. A clear expansion signature is visible well resembling the expected behavior by \citet{roman-zuniga19}.
    The typical arrow size corresponds to a velocity of $\sim1.5$ km/s. 
    {\bf Panels b}: radial and tangential velocities as a function of the distance from the system barycenter. 
    The positions of the identified clusters are highlighted with blue arrows.}
    \label{fig:map}
\end{figure}

\section{3D kinematical properties and structure}

\subsection{Kinematical properties}
The mean motion of each star cluster in the plane of the sky has been obtained by using Gaia-EDR3 
proper motions of stars within confidence radii in the range of $16.5\arcmin-25\arcmin$ from each clusters' center (Table~1) 
and by adopting Equation (2) from \citet{gaia18}. Figure 1a shows the derived cluster velocity vectors with 
respect to the entire system mean motion. All velocities were corrected for the effects of perspective 
expansion/contraction by using the procedures described in \citet{vdv06}, which however, result to be negligible
 (of the order of few cents of km/s) with respect to the typical stellar motions. By construction, all
  clusters' average motions reciprocally coincide within $\sim3$ km/s in the sky component of the velocity vector. 
We then split the entire system in a $20\times20$ regular cell grid with each cell having a side of $16.5\arcmin$
and including only stars with G$<16$. In each cell, we derived the average motions of stars with 
respect to the entire system mean motion. Also in this case, we properly accounted for the 
effects of perspective expansion/contraction. The resulting velocity map is shown in Figure 4a, 
where an almost regular expansion pattern is clearly visible. 
Such a kinematical behavior might be due to a combination of effects associated with the dynamics of the 
violent relaxation phase and the dynamical response of the cluster to the mass loss from supernovae explosions 
and gas expulsion, all of which can cause a significant cluster radial expansion. Indeed, a significant fraction 
of nearby young ($t < 5$ Myr) clusters and associations has been recently found to show a similar behavior based 
on Gaia DR2 data \citep{kuhn19}. We stress here that the expansion of a stellar system is a transient process 
not associated to a single evolutionary path and eventual fate. We further discuss this point in Section 4.2.
We also note that in the specific case of the analyzed region this behavior can be in part linked to the so-called ``Hubble Flow-like'' 
large scale motion that affects 
regions extending for several tens of degrees in the Perseus Arm and 
which has been observed \citep{roman-zuniga19} to be moving the Perseus Arm away from the W3-W4-W5 massive star 
formation zones with a velocity of 15 km/s/kpc. 
 
The panels in Figure 4b show the smoothed density distributions of both the radial and tangential
velocities as a function of the distance from the system barycenter. 
The radial component shows a quite narrow distribution slowly increasing as a 
function of the distance and reaching a peak of $\sim2$ km/s at $R\sim100\arcmin$, thus confirming the expansion signal. 
The tangential velocity component shows a broader distribution, with a slowly declining trend
as a function of the distance that can be indicative of a possible large 
scale (of the order of some degrees) rotation pattern with amplitude of $2-3$ km/s 
(corresponding to about $0.2-0.3$ mas/yr).

\begin{figure}
\centering
	\includegraphics[width=12cm]{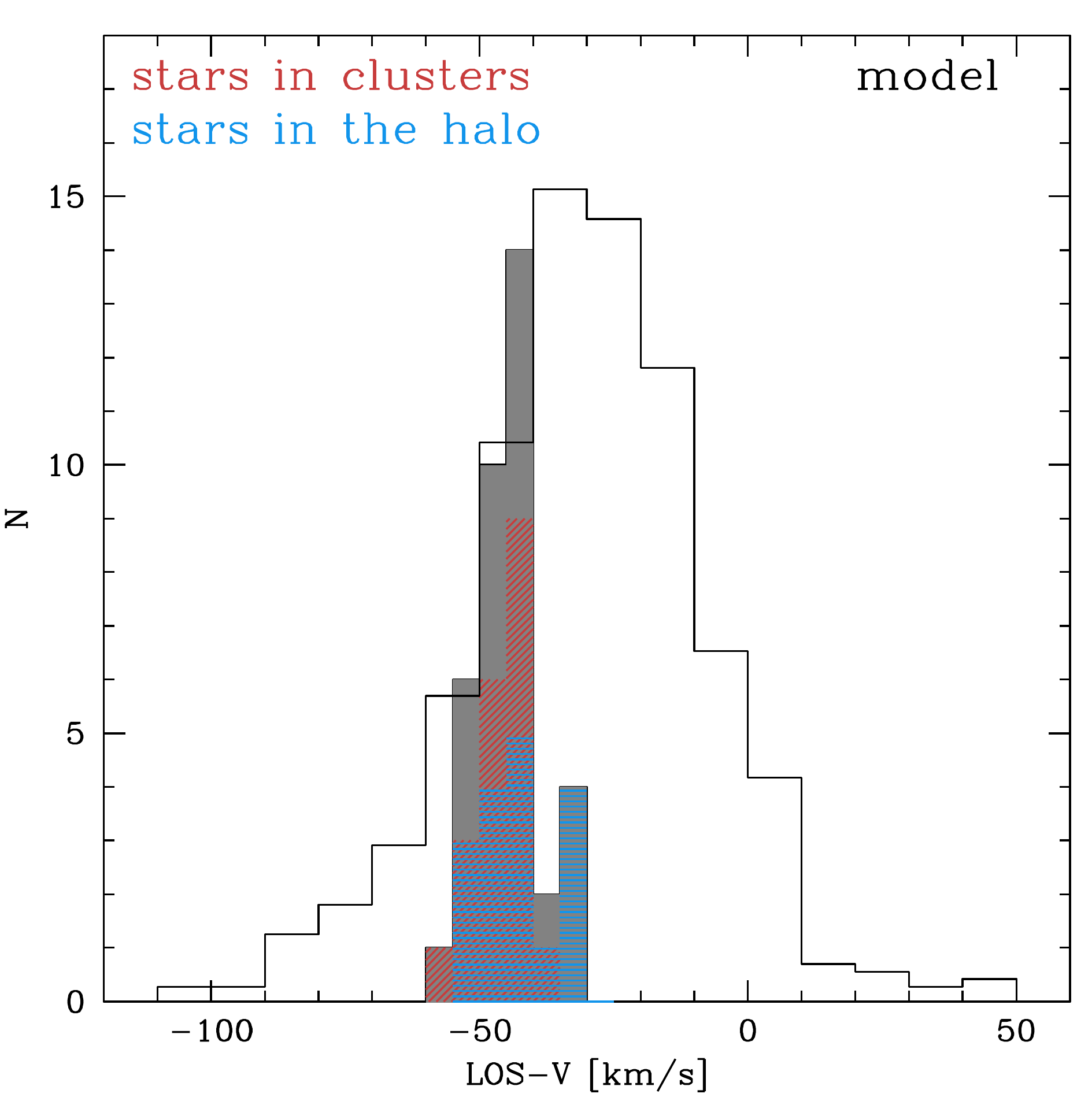}
    \caption{Line of sight velocity distribution of the entire sample of stars observed in LISCA I by using HARPS-TNG 
    as part of the SPA-TNG Large Program. In red is the velocity distribution of stars associated with clusters 
    (see the main text for the selection details) and in blue the distribution of the stars associated with the halo.  
    The grey histogram represents the LOS-V distribution of all the analyzed stars and the black one represents the 
    result of the Besan\c{c}on model.}
    \label{fig:map}
\end{figure}

To check for the line-of-sight velocity (LOS-V) distribution, we complemented Gaia proper 
motions with LOS-Vs obtained with HARPS-N within the SPA-TNG Large Program for the
37 bright stars fulfilling the proper motions and parallax selection 
criteria described before (light blue symbols in Figure 1a; see Section~3).
Accurate (at better than 1 km/s) heliocentric radial velocities for the observed stars 
have been obtained by means of standard cross-correlation technique \citep{tonry79} as implemented in 
the \texttt{IRAF}\footnote{IRAF is distributed by the National Optical Astronomy
Observatories, which is operated by the Association of Universities for
Research in Astronomy, Inc. (AURA) under cooperative agreement with the National
Science Foundation.} \citep{tody86} task \texttt{FXCOR} and suitable synthetic templates 
computed with the \texttt{SYNTHE} code \citep{kurucz73}. 
The median line-of-sight velocity of all the analyzed stars is $(-43.9 \pm 3.5)$ km/s. 

The resulting LOS-V distribution of cluster and halo stars is shown in Figure 5.  
For each cluster, we obtained the mean velocity by averaging out the LOS-Vs of 
stars falling within $16.5\arcmin-25\arcmin$ from the clusters' center. Our analysis shows that 
the clusters are moving at an average LOS-V$_{clusters} \sim-47.4$ km/s with $\sigma\sim2.9$ km/s. 
The stars (17) not directly associable with clusters are here tentatively assigned to the halo. 
The agreement between the line-of-sight motion of such stars and that of the cluster 
is satisfactory, as shown in Figure 5, with an average velocity LOS-V$_{halo} \sim-43.4$ km/s and  $\sigma\sim6.3$ km/s. 
Figure 5 also shows a comparison with the LOS-V distribution obtained from the Besan\c{c}on 
model for the same sample of stars used before. The observed LOS-V distribution is significantly narrower 
than the one expected for field disk stars within the same field of view, thus making it unlikely 
to be just a random extraction of disk star velocities in the area.

\subsection{Mass and structural properties}
A detailed characterization of the individual clusters in the region will be 
the subject of a related paper (E. Dalessandro, in prep.). Here we provide further details on the diffuse stellar halo. 
To determine the projected density profile of the halo we used stars with G$<16$ and 
selected in proper motions and parallaxes as described before. We also subtracted the 
contribution of each cluster by removing all stars located at a distance $r<25\arcmin$ from each cluster center. 
We divided the field of view in 16 concentric annuli centered on the system center and split in a 
number (from 2 to 4 depending on the azimuthal coverage) of sub-sectors. 
We then counted the number of stars lying within each sub-sector and divided 
it by the sub-sector area. The stellar density in each annulus was finally obtained as 
the average of the sub-sector densities, and the standard deviation among the sub-sectors 
densities was adopted as error.
The resulting density profile is shown in Figure 6. The background corrected profile
is characterized by a typical profile with a high-density core and low-density halo that 
can be well fit by a single-mass King model \citep{king66} thus suggesting that the diffuse stellar halo
resembles a low-density cluster-like stellar structure.  
To determine the physical parameters of the halo we used single-mass, spherical and isotropic King models (King 1966). 
This is a single-parameter family, where the shape of the density profile is uniquely determined by the dimensionless 
parameter W$_0$ (or concentration c). To determine the best-fit model we compared the background-subtracted 
surface density profile with the King model family, and found the halo profile to be best-fit by a 
low-density model ($c=0.84$) with large core and half-mass
radii $r_c=1770.0^{+182.4}_{-194.7}\arcsec$ and $R_h=3381.1^{+287.1}_{-304.5}\arcsec$.

We analyzed the mass function (MF) of the likely halo stars. Stellar masses were derived by interpolation as a
function of the G-band magnitude with the 20 Myr isochrone used in Figure 3. 
Starting from the observed 
MF we derived the stellar mass of the entire system.
We normalized both a Salpeter \citet{salpeter55} and a Kroupa \citep{kroupa01} 
theoretical initial mass functions (IMFs)
to the high-mass portion (roughly corresponding to G$<12$) of the observed MF, 
where the photometric completeness of Gaia is expected to be close to $100\%$ 
in relatively low-density environments like those surrounding {\it h} and $\chi$ Persei. 
Then, the total mass is simply given by the integral of the normalized IMF in the mass range 
$(0.1<m<8) M_{\odot}$.   
We find that the total mass ranges from $M_{tot}\sim6\times10^5 M_{\odot}$ when a Salpeter IMF is adopted to 
$1.5\times10^5 M_{\odot}$ when a Kroupa IMF is used instead, making it compatible with the mass 
regime of the so called Young Massive Clusters (YMCs) 
and with the peak of the present-day mass distribution of Galactic globular clusters. 
We find that the mass is equally split between clusters and the diffuse stellar halo. 
Stellar clusters have masses ranging from $\sim2\times10^4 M_{\odot}$ as for $\it h$ and $\chi$ Persei 
(in quite good agreement with estimates from \citealt{currie10}) and $\sim 500 M_{\odot}$ for N3.
By adopting the same heliocentric 
distance as before and a galactocentric distance of 9.7 kpc, 
we find that the formal Jacobi radius of the system (which gives an approximate
indication of the expected possible extension of the system), 
ranges from $\sim140\arcmin$ to $\sim200\arcmin$ (corresponding to 91-132 pc) depending on the adopted mass, 
thus nicely matching the observed properties of the analyzed structure.

We emphasize that the observed radial variation of the LISCA I halo 
is not expected for a generic sample of field stars. The radial profile we found thus provide further evidence that 
the halo identified in our analysis is one of the dynamical components of the complex cluster structure.

In addition, by combining all the structural and kinematic information collected so far we can calculate both the 
one dimensional 1D ($\sigma_{1D}$)
and the virial velocity dispersions ($\sigma_{vir}^2=GM/\eta R_h$, see e.g. equation 19 in \citealt{kuhn19}), 
where $\sigma_{vir}$
represents the velocity dispersion of a virialized stellar system. 
\citet{kuhn19} used these two quantities for an approximate determination of the dynamical state of 
28 young (1-5 Myr) nearby clusters.
Following \citet{kuhn19}, we have adopted $\eta\sim10$ as appropriate
for a stellar system with a Plummer star density distribution.  
We find that $\sigma_{1D}=1.54 \pm 0.06$ km/s and $\sigma_{vir}$ ranges from $1.29$ km/s and $2.63$ km/s 
for $R_h=3381.1\arcsec$ and assuming a total mass in the $1.5-6 \times 10^5 M_{\odot}$ range.
While these values provide only an approximate description of the system's dynamical state, 
they suggest it should be gravitationally bound.
   
The purely observational results collected so far show that 
a) all stars and clusters distributed close to {\it h} and $\chi$ Persei define a quite complex 
structure with evidence of possible mutual interactions in the form of intra-cluster 
over-densities and/or bridges, b) they have the same metallicity,  $\alpha$-element abundance 
and age within a small interval, c) the stellar density distribution 
of the surrounding diffuse stellar halo resembles that of a cluster-like stellar 
system. 
The combination of these evidences clearly suggests that stars distributed within 
a few degrees from {\it h} and $\chi$ Persei are part of a common, sub-structured and recently
 formed/still forming stellar system that we name hereafter LISCA I\footnote{LISCA stands for Lively Infancy of Star Clusters and
 Associations }.

\begin{figure}
\centering
	\includegraphics[width=12cm]{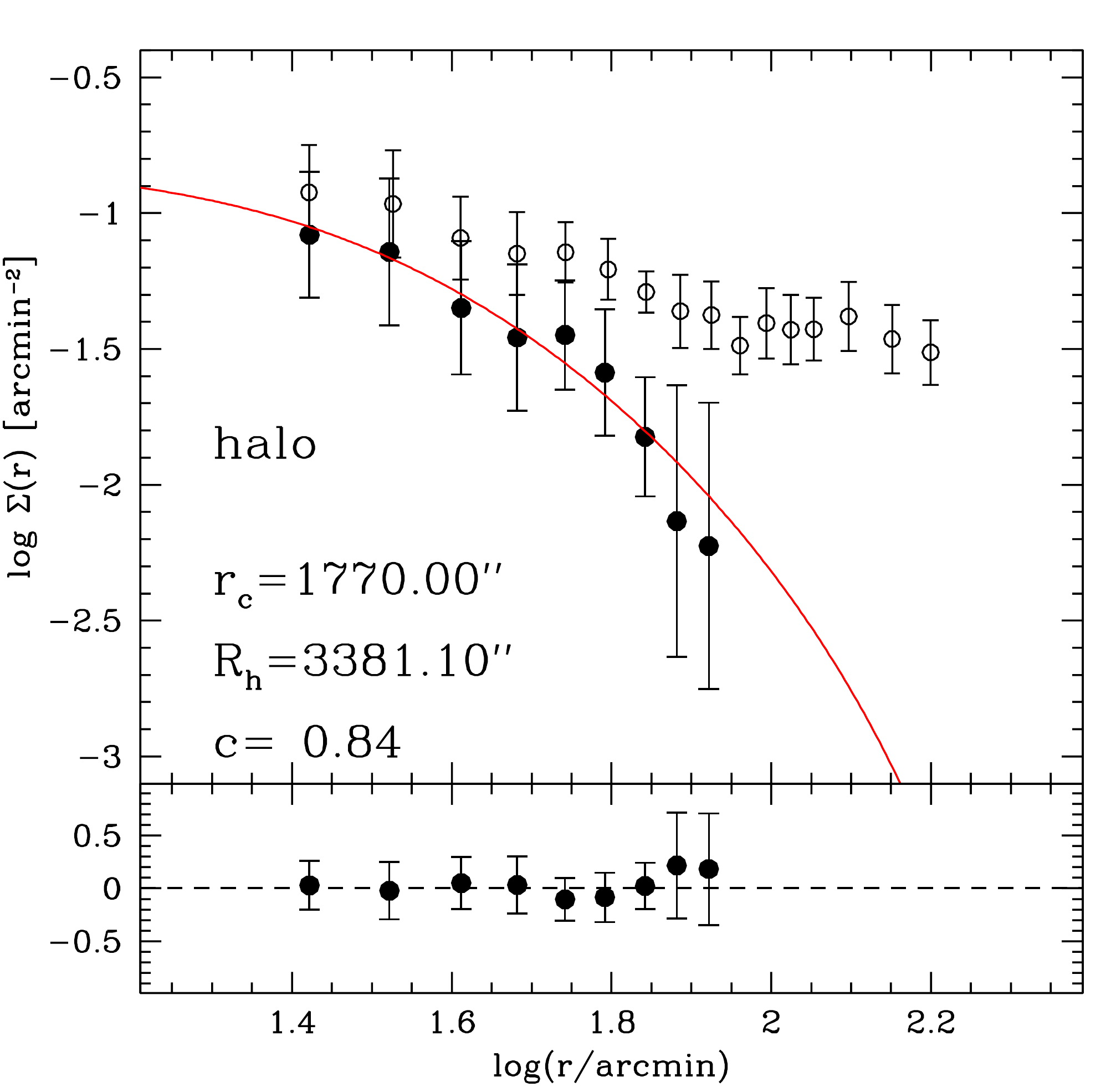}
    \caption{Density profile of the diffuse stellar halo obtained within the selected 
    field of view by subtracting the contribution of clusters as described in the text. 
    Open circles represent the observed density, while the black ones are the background subtracted density values. 
    In red, the best fit mono-mass King model is super-imposed to the observed values.}
    \label{fig:map}
\end{figure}

\begin{figure}
	\includegraphics[width=\columnwidth]{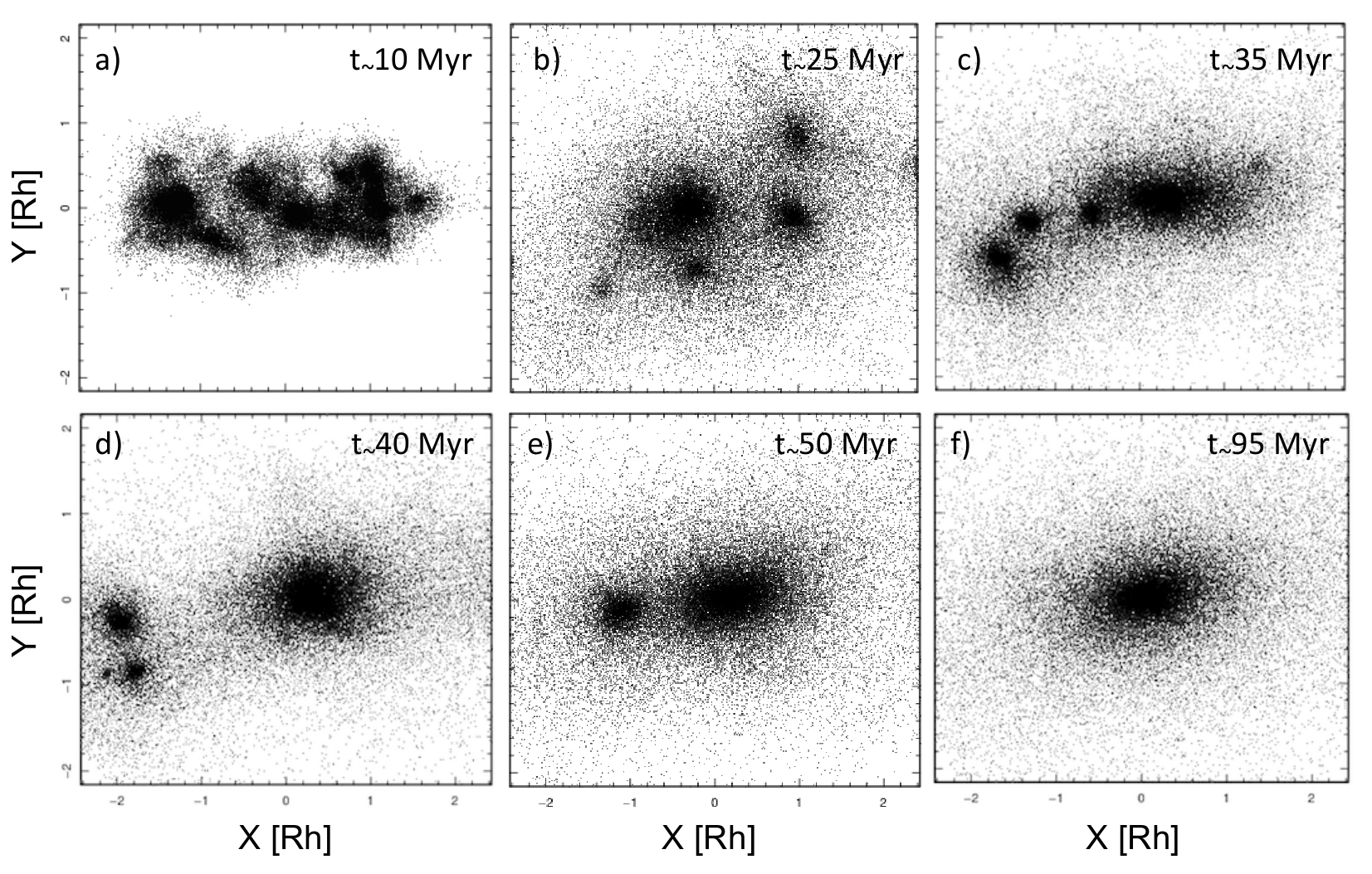}
    \caption{Schematic view of a hierarchical cluster formation path as traced by the 
    $N$-body simulations presented in this work. 
    The axis are expressed in unit of the system Rh.}
    \label{fig:map}
\end{figure}

\section{N-body simulations of a LISCA I - like system}
Several theoretical studies have explored the early dynamical evolution of star clusters and tried to establish 
a  connection between the initial structural and kinematical properties of molecular clouds to the formation of single 
and multiple stellar clusters (see for example \citealt{goodwin04,allison10,parker14,banerjee14,banerjee15,
fujii16,arnold17,sills18}).
In order to provide an example of the evolutionary history behind a system with structural 
and kinematical properties similar to those found in LISCA I, we show here the results of one 
of the realisations of a large suite of direct $N$-body simulations aimed 
at investigating the violent relaxation phase of a cluster, starting from initial conditions
 characterized by non-vanishing total angular momentum and a fractal density distribution, 
 to heuristically represent a turbulent giant molecular cloud in a variety of kinematic and structural initial states.

While a comprehensive description and phase space analysis of this suite of $N$-body simulations along with a comparison with
previous works will be presented in a separate article (A.L. Varri et al., in preparation), 
here we provide a short summary of the key properties of the survey. 
The initial conditions for these simulations were generated by using \texttt{McLuster} \citep{kuepper11}, 
with both homogeneous and inhomogeneous mass distributions.  
The values of the fractal dimension explored in the survey range from 1.6 to 3.0, 
with D=3.0 corresponding to a homogeneous stellar system.  As for the stellar velocities,
the systems have been initialized with a variety of ``temperatures'' of the stellar distribution,
with a ratio of the kinetic energy due to random stellar motions to the potential
energy of the system ranging from 0.1 (`cold') to 0.25 (`warm'). 
With these initial conditions, the system is typically initially out of virial equilibrium,
undergoes the so-called violent relaxation process \citep{aarseth88}, and eventually settles into
an equilibrium configuration in a few dynamical times. Such a process naturally produces
a characteristic pressure anisotropy signature \citep{vanalbada82,trenti05,vesperini14} 
in the velocity space of the resulting configuration.
In addition, we have also explored the effects of the presence of some primordial angular momentum
\citep{gott73,boily99}.
Internal rotation was added to the initial conditions such that the ratio between
the kinetic energy due to ordered motions and the potential energy could assume a range
of values from 0.0 (non-rotating) to 0.5 (moderately rotating), 
and the initial rotation curve is always approximately solid-body.
All $N$-body simulations have N=65536 equal-mass particles and the simulations have been performed
with the direct summation code \texttt{Starlab} \citep{PZ01}. The initial mass distribution of the simulation depicted
in Figure~7 is characterized by a value of the fractal dimension parameter D=2.4. 
Such a definition corresponds to significant deviations from a uniform spatial distribution
of stellar positions. The initial kinematic state is `cold' (random kinetic energy/potential energy = 0.1)
and moderately rotating (ordered kinetic energy/potential energy =0.5). 

A representative example of the early dynamical evolution explored in such $N$-body simulations
is illustrated in Figure 7 and can be summarized as follows:
a) at early stages the distribution of stars preserves the initial inhomogeneities and tens 
of small clumps ($\sim100 M_{\odot}$) can be identified (Figure 7a). 
Several of these clumps are then destroyed because of dynamical interactions 
and some of their stars contribute to the surrounding field and to the formation
 of a gravitationally bound halo (Figures 7b,c);
b) At later time, the surviving clumps merge to form a few more massive ($10^3 - 10^4 M_{\odot}$)
 and larger clusters (Figure 7d,e);
c) Finally, one or two clusters survive this hierarchical merger process and will 
eventually evolve into a single massive cluster (Figure 7f).

\begin{figure}
\centering
	\includegraphics[width=12cm]{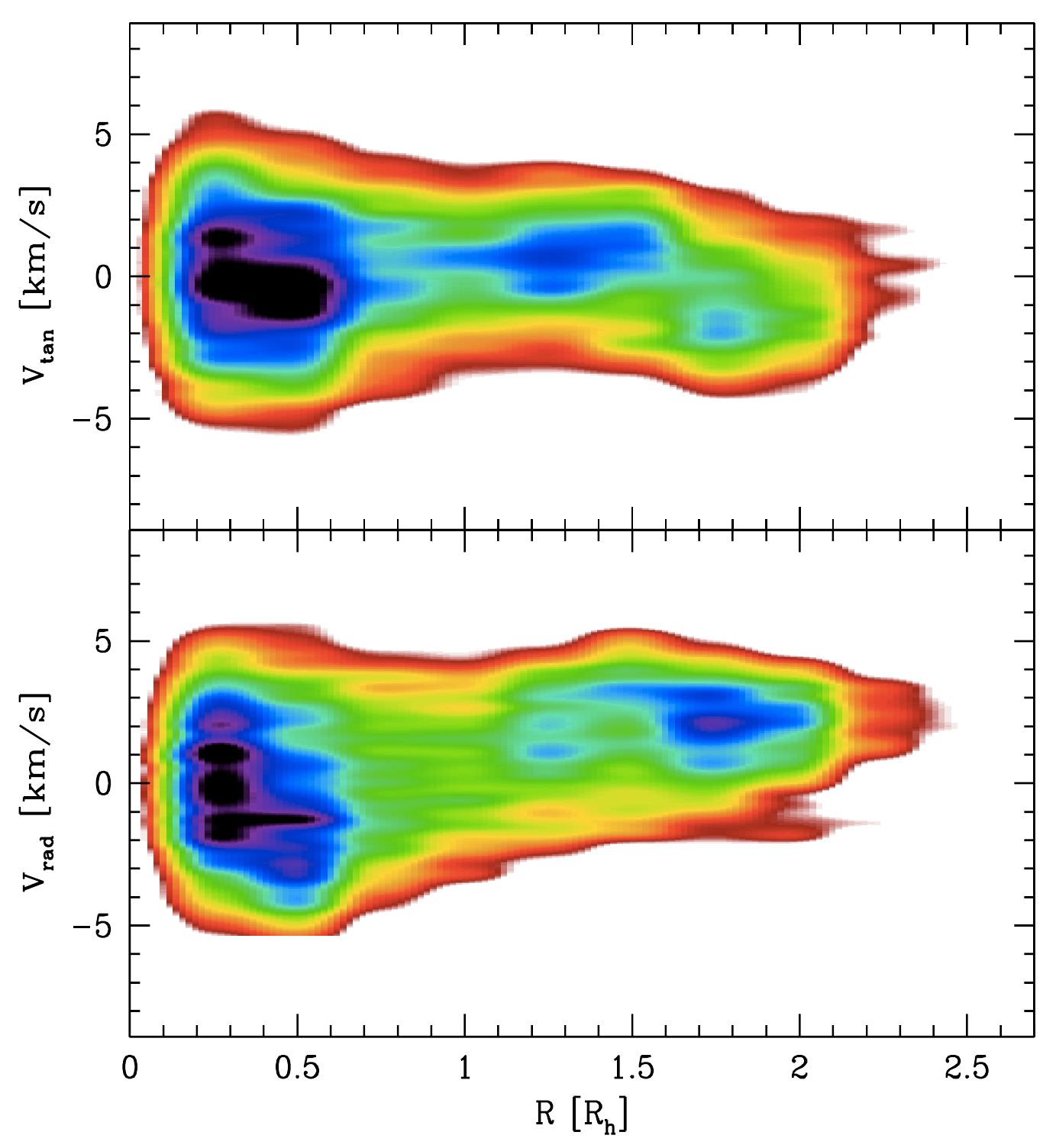}
    \caption{Radial and tangential velocity distributions for the simulation snapshot corresponding to Figure 7c. }
    \label{fig:map}
\end{figure}

We emphasize that the snapshots shown here are not meant to provide a dynamical model
tailored to the evolution of the LISCA I system, but just to illustrate the typical
dynamical evolution of a hierarchical cluster assembly process in the presence of a 
realistic degree of kinematic complexity. Such complexity, which is often neglected
in star cluster formation studies, has a key role in determining the properties of
the violent relaxation phase, the longevity of the population of sub-clumps, and the structural 
and kinematic properties of the resulting end-product massive cluster. 
The six snapshots in Figure 7 show the system after about 1.4, 3.0, 4.6, 5.4, 6.7, and 
12.6 free-fall timescales ($t_{ff}$). Assuming an initial mass of $M_i\sim6\times10^5 M_{\odot}$ and 
a 3D half-mass radius $r_h\sim40$pc, the snapshot in Figure 7b would represent 
the structure of the system at the approximate age of the LISCA I system ($20-25$ Myr).
Therefore, the observational properties of {\it h} and $\chi$ Persei and their surrounding structures,
fit well within an early stage of the hierarchical cluster assembly process (Figures 7b,c), 
which may potentially lead the system to the formation of a stellar system of 
some $10^5 M_{\odot}$ within a short timescale ($\sim100$ Myr -- Figure 7f). 

Also, the kinematical patterns corresponding to the early snapshots of Figure 7 (see Figure 8) 
nicely resemble those observed in LISCA I. In particular, we note that also the expansion pattern observed in LISCA I 
is compatible with the results from our suite of simulations and shown in the lower panel of Figure 8 to compare with Figure 4b. 
As discussed in Section~4.1, although there might be different physical mechanisms responsible for the expansion of the system,
the (qualitative) agreement with the simulations suggests that it is compatible with the dynamical evolution 
of a system in the early phases of a hierarchical cluster assembly. 
 
Finally we point out that although our simulations explore the evolution of systems starting 
with a clumpy mass distribution similar to one that could arise from a hierarchical star formation process, 
structural properties like those observed in the current dynamical phase of LISCA I could also result 
from the growth of density fluctuations in a system emerging with a more homogeneous initial mass 
distribution from a monolithic formation event (see e.g. \citealt{aarseth88}).

\section{Summary and Conclusions}
The results presented in this paper suggest that we might have caught in the act, for the first time in our ``courtyard'',
an ongoing hierarchical cluster assembly process for which the structural and kinematical properties of the involved 
components are described in detail. 

It is hard to make strong predictions about the final by-product of LISCA I
and the fraction of mass that will be
actually retained during its evolution, as it may depend on many factors including the nature and duration of the observed 
expansion. 
However, the results obtained here suggest that the formation of small stellar 
structures and their subsequent growth driven by dynamical interactions might have strongly contributed 
to shape the observed properties of LISCA I, thus possibly representing
a viable process to form massive and long-lived stellar systems also in relatively low-density environments.
This mechanism was long believed to work only in massive starburst galaxies, 
such as M83 \citep{bastian11} and M51 \citep{chandar11}, but to be inefficient in local, 
lower density disk galaxies such as the Milky Way 
or M31 \citep{krumholz12}.  

In case LISCA I will be able to continue its evolution as schematically shown in Figure 7, it 
can produce a YMC or alternatively, given its extension ($R_h\sim35$pc), 
it could also be identified as the progenitor of a 
so-called diffuse star cluster (also known as ``faint fuzzy"; \citealt{larsen00}), 
that is characterized by 
a mass comparable to that of globular clusters while being significantly more extended. 
These systems have a strongly debated origin \citep{fellhauer02}, they are observed in external galaxies \citep{huxor05}, 
but have remained elusive in the Milky Way so far. 
In any case, these results may give a boost 
to the possibility of studying massive cluster formation 
by resolving individual stars in our neighborhood thus constraining 
the physical mechanisms with a level of detail that cannot be achieved for distant systems. 
Interestingly in this respect, the properties of the interstellar medium in the Galactic 
disk are similar to those in the disks of other nearby galaxies \citep{longmore13}. As a consequence, 
our understanding of star formation and feedback in close young clusters and associations
 can probe star formation across much of the local Universe. 
 
Recent results obtained by Gaia \citep{cantat-gaudin18,castro-ginard18} 
have shown that the Galactic disk structure is 
composed by large numbers of stellar groups, hierarchies, young clusters with most
of them being distributed along filamentary, string-like structures with similar
properties to those observed in the Perseus Arm and in the LISCA I (see for example \citealt{kuhn19}).
This opens the possibility to check whether the hierarchical assembly process framed here can be efficient 
within disk-like galaxies such as the Milky Way.
Our study would therefore introduce a possible new picture in which the different
appearance and properties of young stellar aggregates from clumps and filaments
as observed in the Galaxy spiral arms (e.g. \citealt{kounkel19,kuhn19}), to double/multiple clusters as found in 
the Magellanic Cloud disks (e.g. \citealt{dieball02,mucciarelli12,dalessandro18}) and finally to massive globulars, can be interpreted as 
the morphological evidence of different evolutionary phases of the same hierarchical assembly process. 
This has important implications on our understanding of the environmental conditions 
(both locally and in the distant Universe) necessary to form massive stellar clusters and,
 as consequence, on the evolution of stellar cluster properties over cosmic time. 
 In this respect, it is extremely interesting to note that the hierarchical build-up 
 of massive stellar clusters may play an important role during the formation of the multiple 
 stellar populations observed in old GCs and could be one of the key ingredients necessary 
 to explain the variety of stellar populations' properties \citep{bekki17,bailin18,howard19}.
Finally, the results obtained for LISCA I and the $N$-body simulations also show that 
the kinematic state of the initial molecular cloud play a key role in prolonging the 
survival timescales of sparse and low-mass clumps and associations, thus mitigating 
the discrepancy between their observed and expected numbers in the Galaxy and providing 
further support to a primordial origin of the `kinematic complexity' currently emerging in old GCs 
\citep{lanzoni18,kamann18}.

\software{DBSCAN, FXCOR \citep{tonry79}, GOFIO \citep{rainer18}, IRAF \citep{tody86}, 
MARCS \citep{gustafsson08}, McLuster \citep{kuepper11}, 
Starlab \citep{PZ01}, SYNTHE \citep{kurucz73}, Turbospectrum \citep{alvarez98,plez12}}

\section*{Acknowledgements}
The authors thank the anonymous referee for the careful
reading of the paper and the useful comments and suggestions.
ED and LO acknowledge financial support 
from the project {\it Light-on-Dark} granted by MIUR through
PRIN2017-2017K7REXT contract and from the Mainstream project {\it SC3K - Star clusters in the inner 3 kpc} (1.05.01.86.21)
granted by INAF.
ALV acknowledges support from a UKRI Future Leaders Fellowship (MR/S018859/1). 
MB acknowledges the financial support to this research by INAF, through the Mainstream Grant 1.05.01.86.22 assigned to 
the project {\it Chemo-dynamics of globular clusters: the Gaia revolution}.
SS gratefully acknowledges financial support from the European Research Council (ERC-CoG-646928, Multi-Pop).
MF acknowledges the financial support of the Istituto Nazionale di Astrofisica (INAF), Osservatorio Astronomico di Roma, 
and Agenzia Spaziale Italiana (ASI) under contract to INAF: ASI 2014-049-R.0 dedicated to SSDC.
This work has made use of data from the European Space Agency (ESA)
mission {\it Gaia} (\url{https://www.cosmos.esa.int/gaia}), processed by
the {\it Gaia} Data Processing and Analysis Consortium (DPAC,
\url{https://www.cosmos.esa.int/web/gaia/dpac/consortium}). Funding
for the DPAC has been provided by national institutions, in particular
the institutions participating in the {\it Gaia} Multilateral Agreement.





\begin{thebibliography}{99}

\bibitem[Aarseth et al.(1988)]{aarseth88} Aarseth, S.~J., Lin, D.~N.~C., \& Papaloizou, J.~C.~B.\ 1988, \apj, 324, 288

\bibitem[Adamo et al.(2015)]{adamo15} Adamo, A., Kruijssen, J.~M.~D., Bastian, N., et al.\ 2015, \mnras, 452, 246

\bibitem[Adamo et al.(2020)]{adamo20} Adamo, A., Hollyhead, K., Messa, M., et al.\ 2020, \mnras, doi:10.1093/mnras/staa2380

\bibitem[Allison et al.(2010)]{allison10} Allison, R.~J., Goodwin, S.~P., Parker, R.~J., et al.\ 2010, \mnras, 407, 1098

\bibitem[Alvarez \& Plez(1998)]{alvarez98} Alvarez, R. \& Plez, B.\ 1998, \aap, 330, 1109

\bibitem[Arnold et al.(2017)]{arnold17} Arnold, B., Goodwin, S.~P., Griffiths, D.~W., et al.\ 2017, \mnras, 471, 2498

\bibitem[Banerjee \& Kroupa(2014)]{banerjee14} Banerjee, S. \& Kroupa, P.\ 2014, \apj, 787, 158

\bibitem[Banerjee \& Kroupa(2015)]{banerjee15} Banerjee, S. \& Kroupa, P.\ 2015, \mnras, 447, 728

\bibitem[Banerjee \& Kroupa(2017)]{banerjee17} Banerjee, S. \& Kroupa, P.\ 2017, \aap, 597, A28

\bibitem[Banerjee \& Kroupa(2018)]{banerjee18} Banerjee, S. \& Kroupa, P.\ 2018, The Birth of Star Clusters, 143

\bibitem[Baumgardt \& Kroupa(2007)]{baumgardt07} Baumgardt, H. \& Kroupa, P.\ 2007, \mnras, 380, 1589

\bibitem[Bailin(2018)]{bailin18} Bailin, J.\ 2018, \apj, 863, 99

\bibitem[Bastian et al.(2011)]{bastian11} Bastian, N., Adamo, A., Gieles, M., et al.\ 2011, \mnras, 417, L6

\bibitem[Bastian \& Lardo(2018)]{bastian18} Bastian, N. \& Lardo, C.\ 2018, \araa, 56, 83

\bibitem[Beccari et al.(2018)]{beccari18} Beccari, G., Boffin, H.~M.~J., Jerabkova, T., et al.\ 2018, \mnras, 481, L11. doi:10.1093/mnrasl/sly144

\bibitem[Bekki(2017)]{bekki17} Bekki, K.\ 2017, \mnras, 469, 2933

\bibitem[Boily et al.(1999)]{boily99} Boily, C.~M., Clarke, C.~J., \& Murray, S.~D.\ 1999, \mnras, 302, 399

\bibitem[Bonnell et al.(2003)]{bonnell03} Bonnell, I.~A., Bate, M.~R., \& Vine, S.~G.\ 2003, \mnras, 343, 413

\bibitem[Bressan et al.(2012)]{bressan12} Bressan, A., Marigo, P., Girardi, L., et al.\ 2012, \mnras, 427, 127

\bibitem[Brodie \& Strader(2006)]{brodie06} Brodie, J.~P. \& Strader, J.\ 2006, \araa, 44, 193

\bibitem[Cantat-Gaudin et al.(2018)]{cantat-gaudin18} Cantat-Gaudin, T., Jordi, C., Vallenari, A., et al.\ 2018, \aap, 618, A93

\bibitem[Carpenter et al.(2000)]{carpenter00} Carpenter, J.~M., Heyer, M.~H., \& Snell, R.~L.\ 2000, \apjs, 130, 381

\bibitem[Castro-Ginard et al.(2018)]{castro-ginard18} Castro-Ginard, A., Jordi, C., Luri, X., et al.\ 2018, \aap, 618, A59

\bibitem[Chandar et al.(2011)]{chandar11} Chandar, R., Whitmore, B.~C., Calzetti, D., et al.\ 2011, \apj, 727, 88

\bibitem[Cosentino et al.(2014)]{cosentino14} Cosentino, R., Lovis, C., Pepe, F., et al.\ 2014, \procspie, 9147, 91478C

\bibitem[Currie et al.(2010)]{currie10} Currie, T., Hernandez, J., Irwin, J., et al.\ 2010, \apjs, 186, 191

\bibitem[Dalessandro et al.(2012)]{dalessandro12} Dalessandro, E., Schiavon, R.~P., Rood, R.~T., et al.\ 2012, \aj, 144, 126

\bibitem[Dalessandro et al.(2015)]{dalessandro15} Dalessandro, E., Miocchi, P., Carraro, G., et al.\ 2015, \mnras, 449, 1811

\bibitem[Dalessandro et al.(2018)]{dalessandro18} Dalessandro, E., Zocchi, A., Varri, A.~L., et al.\ 2018, \mnras, 474, 2277. doi:10.1093/mnras/stx2892

\bibitem[Di Carlo et al.(2019)]{dicarlo19} Di Carlo, U.~N., Giacobbo, N., Mapelli, M., et al.\ 2019, \mnras, 487, 2947

\bibitem[Dieball et al.(2002)]{dieball02} Dieball, A., M{\"u}ller, H., \& Grebel, E.~K.\ 2002, \aap, 391, 547. doi:10.1051/0004-6361:20020815

\bibitem[Elmegreen \& Hunter(2010)]{elmegreen10} Elmegreen, B.~G. \& Hunter, D.~A.\ 2010, \apj, 712, 604

\bibitem[Fellhauer \& Kroupa(2002)]{fellhauer02} Fellhauer, M. \& Kroupa, P.\ 2002, \aj, 124, 2006

\bibitem[Forbes et al.(2018)]{forbes18} Forbes, D.~A., Bastian, N., Gieles, M., et al.\ 2018, Proceedings of the Royal Society of London Series A, 474, 20170616

\bibitem[Fujii \& Portegies Zwart(2016)]{fujii16} Fujii, M.~S. \& Portegies Zwart, S.\ 2016, \apj, 817, 4

\bibitem[Gaia Collaboration et al.(2018)]{gaia18} Gaia Collaboration, Brown, A.~G.~A., Vallenari, A., et al.\ 2018, \aap, 616, A1

\bibitem[Gaia Collaboration et al.(2020)]{gaia20} Gaia Collaboration, Brown, A.~G.~A., Vallenari, A., et al.\ 2020, arXiv:2012.01533

\bibitem[Gieles et al.(2006)]{gieles06} Gieles, M., Portegies Zwart, S.~F., Baumgardt, H., et al.\ 2006, \mnras, 371, 793

\bibitem[Gieles \& Portegies Zwart(2011)]{gieles11} Gieles, M. \& Portegies Zwart, S.~F.\ 2011, \mnras, 410, L6

\bibitem[Goodwin \& Whitworth(2004)]{goodwin04} Goodwin, S.~P. \& Whitworth, A.~P.\ 2004, \aap, 413, 929

\bibitem[Gott(1973)]{gott73} Gott, R.~J.\ 1973, \apj, 186, 481

\bibitem[Gratton et al.(2019)]{gratton19} Gratton, R., Bragaglia, A., Carretta, E., et al.\ 2019, \aapr, 27, 8

\bibitem[Gustafsson et al.(2008)]{gustafsson08} Gustafsson, B., Edvardsson, B., Eriksson, K., et al.\ 2008, \aap, 486, 951

\bibitem[Hayden et al.(2014)]{hayden14} Hayden, M.~R., Holtzman, J.~A., Bovy, J., et al.\ 2014, \aj, 147, 116

\bibitem[Helmi et al.(2018)]{helmi18} Helmi, A., Babusiaux, C., Koppelman, H.~H., et al.\ 2018, \nat, 563, 85

\bibitem[Howard et al.(2019)]{howard19} Howard, C.~S., Pudritz, R.~E., Sills, A., et al.\ 2019, \mnras, 486, 1146

\bibitem[Huxor et al.(2005)]{huxor05} Huxor, A.~P., Tanvir, N.~R., Irwin, M.~J., et al.\ 2005, \mnras, 360, 1007

\bibitem[Kamann et al.(2018)]{kamann18} Kamann, S., Husser, T.-O., Dreizler, S., et al.\ 2018, \mnras, 473, 5591

\bibitem[King(1966)]{king66} King, I.~R.\ 1966, \aj, 71, 64

\bibitem[Kounkel \& Covey(2019)]{kounkel19} Kounkel, M. \& Covey, K.\ 2019, \aj, 158, 122

\bibitem[Kroupa(2001)]{kroupa01} Kroupa, P.\ 2001, \mnras, 322, 231

\bibitem[Krumholz et al.(2012)]{krumholz12} Krumholz, M.~R., Dekel, A., \& McKee, C.~F.\ 2012, \apj, 745, 69

\bibitem[Krumholz et al.(2019)]{krumholz19} Krumholz, M.~R., McKee, C.~F., \& Bland-Hawthorn, J.\ 2019, \araa, 57, 227

\bibitem[Kuepper et al.(2011)]{kuepper11} Kuepper, A.~H.~W., Maschberger, T., Kroupa, P., et al.\ 2011, Astrophysics Source Code Library

\bibitem[Kuhn et al.(2019)]{kuhn19} Kuhn, M.~A., Hillenbrand, L.~A., Sills, A., et al.\ 2019, \apj, 870, 32

\bibitem[Kuhn et al.(2020)]{kuhn20} Kuhn, M.~A., Hillenbrand, L.~A., Carpenter, J.~M., et al.\ 2020, \apj, 

\bibitem[Kurucz(1973)]{kurucz73} Kurucz, R.~L.\ 1973, Ph.D. Thesis

\bibitem[Lada \& Lada(2003)]{lada03} Lada, C.~J. \& Lada, E.~A.\ 2003, \araa, 41, 57

\bibitem[Lanzoni et al.(2018)]{lanzoni18} Lanzoni, B., Ferraro, F.~R., Mucciarelli, A., et al.\ 2018, \apj, 861, 16

\bibitem[Larsen \& Brodie(2000)]{larsen00} Larsen, S.~S. \& Brodie, J.~P.\ 2000, \aj, 120, 2938

\bibitem[Li et al.(2019)]{li19} Li, C., Sun, W., de Grijs, R., et al.\ 2019, \apj, 876, 65

\bibitem[Lim et al.(2020)]{lim20} Lim, B., Hong, J., Yun, H.-S., et al.\ 2020, \apj, 899, 121. doi:10.3847/1538-4357/aba0a3

\bibitem[Longmore et al.(2013)]{longmore13} Longmore, S.~N., Bally, J., Testi, L., et al.\ 2013, \mnras, 429, 987

\bibitem[Longmore et al.(2014)]{longmore14} Longmore, S.~N., Kruijssen, J.~M.~D., Bastian, N., et al.\ 2014, Protostars and Planets
VI, 291

\bibitem[McKee \& Ostriker(2007)]{mckee07} McKee, C.~F. \& Ostriker, E.~C.\ 2007, \araa, 45, 565

\bibitem[McMillan et al.(2007)]{mcmillan07} McMillan, S.~L.~W., Vesperini, E., \& Portegies Zwart, S.~F.\ 2007, \apjl, 655, L45

\bibitem[Meingast et al.(2019)]{meingast19} Meingast, S., Alves, J., \& F{\"u}rnkranz, V.\ 2019, \aap, 622, L13. doi:10.1051/0004-6361/201834950

\bibitem[Mikolaitis et al.(2014)]{mikolaitis14} Mikolaitis, {\v{S}}., Hill, V., Recio-Blanco, A., et al.\ 2014, \aap, 572, A33

\bibitem[Moeckel \& Bate(2010)]{moeckel10} Moeckel, N. \& Bate, M.~R.\ 2010, \mnras, 404, 721

\bibitem[Mucciarelli et al.(2012)]{mucciarelli12} Mucciarelli, A., Origlia, L., Ferraro, F.~R., et al.\ 2012, \apjl, 746, L19. doi:10.1088/2041-8205/746/2/L19

\bibitem[Oliva et al.(2012)]{oliva12} Oliva, E., Origlia, L., Maiolino, R., et al.\ 2012, \procspie, 8446, 84463T

\bibitem[Origlia et al.(2019)]{origlia19} Origlia, L., Dalessandro, E., Sanna, N., et al.\ 2019, \aap, 629, A117

\bibitem[Parker et al.(2014)]{parker14} Parker, R.~J., Wright, N.~J., Goodwin, S.~P., et al.\ 2014, \mnras, 438, 620

\bibitem[Plez(2012)]{plez12} Plez, B.\ 2012, Astrophysics Source Code Library

\bibitem[Portegies Zwart et al.(2001)]{PZ01} Portegies Zwart, S.~F., McMillan, S.~L.~W., Hut, P., et al.\ 2001, 
The Influence of Binaries on Stellar Population Studies, 371

\bibitem[Portegies Zwart et al.(2010)]{PZ10} Portegies Zwart, S.~F., McMillan, S.~L.~W., \& Gieles, M.\ 2010, \araa, 48, 431

\bibitem[Rainer et al.(2018)]{rainer18} Rainer, M., Harutyunyan, A., Carleo, I., et al.\ 2018, \procspie, 10702, 1070266

\bibitem[Robin et al.(2003)]{robin03} Robin, A.~C., Reyl{\'e}, C., Derri{\`e}re, S., et al.\ 2003, \aap, 409, 523

\bibitem[Rom{\'a}n-Z{\'u}{\~n}iga et al.(2019)]{roman-zuniga19} Rom{\'a}n-Z{\'u}{\~n}iga, C.~G., Roman-Lopes, A., Tapia, M., et al.\ 2019, \apjl, 871, L12

\bibitem[Salpeter(1955)]{salpeter55} Salpeter, E.~E.\ 1955, \apj, 121, 161

\bibitem[Schlafly \& Finkbeiner(2011)]{schlafly11} Schlafly, E.~F. \& Finkbeiner, D.~P.\ 2011, \apj, 737, 103

\bibitem[Schlegel et al.(1998)]{schlegel98} Schlegel, D.~J., Finkbeiner, D.~P., \& Davis, M.\ 1998, \apj, 500, 525

\bibitem[Sills et al.(2018)]{sills18} Sills, A., Rieder, S., Scora, J., et al.\ 2018, \mnras, 477, 1903. doi:10.1093/mnras/sty681

\bibitem[Tody(1986)]{tody86} Tody, D.\ 1986, \procspie, 627, 733. doi:10.1117/12.968154

\bibitem[Tonry \& Davis(1979)]{tonry79} Tonry, J. \& Davis, M.\ 1979, \aj, 84, 1511

\bibitem[Tozzi et al.(2016)]{tozzi16} Tozzi, A., Oliva, E., Iuzzolino, M., et al.\ 2016, \procspie, 9908, 99086C. doi:10.1117/12.2231898

\bibitem[Trenti et al.(2005)]{trenti05} Trenti, M., Bertin, G., \& van Albada, T.~S.\ 2005, \aap, 433, 57

\bibitem[van de Ven et al.(2006)]{vdv06} van de Ven, G., van den Bosch, R.~C.~E., Verolme, E.~K., et al.\ 2006, \aap, 445, 513

\bibitem[van Albada(1982)]{vanalbada82} van Albada, T.~S.\ 1982, \mnras, 201, 939

\bibitem[Vesperini et al.(2014)]{vesperini14} Vesperini, E., Varri, A.~L., McMillan, S.~L.~W., et al.\ 2014, \mnras, 443, L79

\bibitem[Zari et al.(2018)]{zari18} Zari, E., Hashemi, H., Brown, A.~G.~A., et al.\ 2018, \aap, 620, A172. doi:10.1051/0004-6361/201834150

\end{thebibliography}

\bibliographystyle{aa}







\end{document}